\documentclass[iop]{emulateapj}

\usepackage{verbatim}
\usepackage{graphicx}
\usepackage{placeins}

\slugcomment{Accepted to ApJ}
\shorttitle{Cluster mass function}
\shortauthors{Kuznetsova, Hartmann \& Burkert}

\begin{document}

\title{Gravitational Focusing and the Star Cluster Initial Mass Function}

\author{Aleksandra Kuznetsova\altaffilmark{1}, Lee Hartmann\altaffilmark{1}
and Andreas Burkert\altaffilmark{2,3}}

\altaffiltext{1}{Department of Astronomy, University of Michigan, 830 Dennison, 500 Church St., Ann Arbor,
MI 48109-1042, USA}
\altaffiltext{2}{University Observatory Munich, Scheinerstrasse 1, D-81679 Munich, Germany}
\altaffiltext{3}{Max-Planck-Fellow, Max-Planck-Institute for Extraterrestrial Physics, Giessenbachstrasse 1,
85758 Garching, Germany}

\email{kuza@umich.edu, burkert@usm.lmu.de, lhartm@umich.edu}

\newcommand\msun{\rm M_{\odot}}
\newcommand\mdot{\dot{M}}
\newcommand\lsun{\rm L_{\odot}}
\newcommand\msunyr{\rm M_{\odot}\,yr^{-1}}
\newcommand\be{\begin{equation}}
\newcommand\en{\end{equation}}
\newcommand\kms{\rm{\, km \, s^{-1}}}
\newcommand\K{\rm K}
\newcommand\etal{{\rm et al}.\ }
\newcommand\sd{\partial}
\newcommand\mpctwo{\msun \, {\rm pc^{-2}}}
\newcommand\rhoo{\rho_{\circ}}
\newcommand\cmthree{\rm cm^{-3}}
\newcommand\no{n_{\circ}}

\begin{abstract}
We discuss the possibility that gravitational focusing,
is responsible for the power-law mass function of star clusters
$N(\log M) \propto M^{-1}$.  This power law can be produced
asymptotically when the mass accretion rate of an object
depends upon the mass of the accreting body as $\mdot \propto M^2$.
While Bondi-Hoyle-Littleton accretion formally produces this dependence on
mass in a uniform medium, realistic environments are much more complicated.
However, numerical simulations in SPH allowing for sink formation yield
such an asymptotic power-law mass function.  We perform pure N-body simulations to isolate the effects of gravity from those of gas physics and 
to show that clusters naturally result with the power-law mass distribution.
We also consider the physical conditions necessary to produce clusters
on appropriate timescales.  Our results help support the idea that gravitationally-dominated
accretion is the most likely mechanism for producing the cluster mass function.
\end{abstract}


\section{Introduction}

In recent years significant progress has been made on the form of the stellar cluster
IMF \citep[scIMF;][]{zhang99,degrijs03,mccrady07}.
As summarized by \citet{fallchandar12}, clusters in the Milky
Way \citep{lada03}, Magellanic Clouds, M83, M51, and the Antennae exhibit a similar power-law mass function
with $d \log N/ d \log M \equiv \Gamma = -0.9 \pm 0.15$. Others suggest a Schechter-type mass
function is a better fit, with a truncation at high masses \citep{gieles06,bastian08}; however,
all agree on the power-law behavior of $\Gamma \sim -1$ at lower masses.

One explanation for the scIMF has been that it simply reflects the mass distribution of
parent giant molecular clouds \citep[e.g.,][]{elmegreenefremov97}, perhaps with some
accounting for feedback by stellar energy input \citep{fall10}.  However,
the observed mass functions of inner Milky Way giant
molecular clouds (GMCs) yield $\Gamma \sim -0.5$ \citep{williamsmckee97}; the Antennae
also show similarly flat mass functions \citep{wilson03,wei12}, although outer Milky
Way clouds may exhibit $\Gamma \lesssim -1.1$ and LMC clouds $\Gamma \sim -0.7$ \citep{rosolowsky05}.
Moreover, while GMC masses obviously provide an upper limit to cluster masses, it isn't obvious
why the global cloud mass function should be reflected in cluster masses that are one to two orders
of magnitude smaller.  Even if the fraction of dense gas which can form clusters is similar
among molecular clouds \citep[and the evidence for this is mixed; see][]{bastian08,
lada10,burkerthartmann13,battistiheyer14}, the manner in which the dense gas component might be
broken up into individual protocluster clouds is not necessarily the same
\citep[see ][for a model of modified turbulent acccretion]{hennebelle12}.

Long ago, \citet{zinnecker82} showed that if bodies accrete mass at a rate proportional
to the square of their mass, $\mdot \propto M^2$, a population with a narrow initial mass
range will develop a power-law mass distribution which asymptotically approaches $\Gamma = -1$.
Zinnecker pointed out that ``Bondi-Hoyle-Lyttleton'' (BHL) accretion, with a mass accretion rate characterized by
\begin{equation}
\mdot = {4 \pi G^2 M^2 \rhoo \over (c_s^2 + v_{\infty}^2)^{3/2}} \equiv \alpha M^2\,,
\label{eq:bhl}
\end{equation}
\citep[see discussion in][]{edgar04}
has the requisite dependence on the central gravitating
mass $M$.
 
In the standard picture of BHL accretion, a central gravitating mass, $M$, travels through an infinitely large ambient medium $\rhoo$ which has constant sound speed $c_s$ at constant relative velocity $v_{\infty}$. The mass's gravity focuses material into a wake bound to the object from which it accretes material at the rate in Eq. \ref{eq:bhl}. 
But the molecular clouds within which stars and clusters form
exhibit density and velocity fields that are far from
uniform; in addition, the cloud itself is
self-gravitating.  Nevertheless, using isothermal SPH simulations with decaying turbulence,
\citet{ballesteros15} (BP15) demonstrated the development of power-law sink mass functions
with $\Gamma = -1$.  These results suggest that gravitational focusing - which is at the
heart of the BHL accretion process - is able to operate
despite the complex, time-variable environment.

In this contribution, we examine the conditions for which gravitational focusing
can produce star clusters with the required mass function.  We show that observations
of cluster-forming clouds are consistent with the requirements provided the observed
supersonic velocity distributions are understood as being gravitationally-generated
rather than turbulence driven by an external agent.  We suggest that the signature
of the dominance of gravitational over thermal physics is the power-law behavior
of the mass function 
thereby linking the cluster and stellar IMF. 

\section{Numerical Simulations}

The power-law sink mass functions found by
BP15 were produced in
SPH calculations with an initial supersonic velocity field
but allowing the turbulence to decay. To 
minimize the role
played by thermal physics, an isothermal equation
of state was adopted.  If, as suggested by the possible connection with BHL accretion gravity dominates, a similar
result should be found with gravitational
accretion without including any
gas physics.  We therefore explore results using
a pure N-body code.

\subsection{Numerical Setup}
\par To test the scenario of purely gravitational accretion, we use the \emph{ChaNGa} (Charm N-body GrAvity solver) code \citep{jetley08,menon15}. Since \emph{ChaNGa} is a cosmological code, using the ``dark matter" particle implementation ensures that all interactions are purely gravitational, with no gas physics included. With a standard $\Lambda$-CDM cosmology, we choose a set of units in which the code length unit corresponds to $1 $ $\mathrm{kpc}$ of proper distance, the mass unit is $2.22 \times 10^{5}$ $\mathrm{M_{\odot}}$ and the time unit is  $1$ $\mathrm{Gyr}$. For generality, the simulation can be rescaled to representative values. Particles are originally distributed so as to have a uniform density within the chosen geometry. We generate a three dimensional initial turbulent velocity field with the Kolmogorov power law spectrum $P(k) \propto \mathbf{k}^{-11/3}$ with maximum wavenumber $k_{max} = 64$, so that each run is seeded with turbulent velocity fluctuations. The primary function of the initial turbulence is to `stir' up random density fluctuations. 
\par We include both spherical and disk geometries, run with different random seeds. The initial conditions are sub-virial for both cases, with the initial virial parameter, $\alpha_i = 2 |K|/|U| =  5\sigma^2R /GM$, given in Table \ref{tab:runs}, leading to an overall collapse. All runs have particle distributions of total mass $M=1$, total radius $R=1$, and we use $G=1$ in code units; disk runs start off with a thickness, $h=0.1$. We employ the default periodic boundary conditions, with a box size set to $10R$, to avoid interactions with the boundary. Force smoothing to avoid tight binaries is implemented in the form of a softening parameter $\epsilon$, based on \cite{dehnen01}. Table \ref{tab:runs} lists each run and its relevant parameters.  

\begin{deluxetable}{ l l c c r r r r }
\tablecaption{\emph{ChaNGa} runs \label{tab:runs} }
\tablewidth{\columnwidth}
\tablehead{\colhead{run} & \colhead{N} & \colhead{geometry} & \colhead{seed} & \colhead{$\bar{\ell_i}$}& \colhead{$\epsilon$} & \colhead{min $\ell_{k}$ } & \colhead{$\alpha_{i}$} }
\startdata
40s1b  & 40000  & sphere  & 1 & 0.047 & 0.001 & 0.016 & 0.01 \\                 
40s2b  & 40000  & sphere  & 2 & 0.047 & 0.001 & 0.016 & 0.01 \\                   
40d1b  & 40000  & disk    & 1 & 0.031 & 0.001 & 0.016 & 0.01    \\                    
40d2b  & 40000  & disk    & 2 & 0.031 & 0.001 & 0.016 & 0.01    \\                  
400s1b & 400000 & sphere  & 1 & 0.021 & 7E-4  & 0.016 & 0.01   \\                
400s2b & 400000 & sphere  & 2 & 0.021 & 7E-4  & 0.016 & 0.01 \\
400d1b & 400000 & disk    & 1 & 0.015 & 7E-4  & 0.016 & 0.01 \\                       
400d2b & 400000 & disk    & 2 & 0.015 & 7E-4  & 0.016 & 0.01 \\                    
\enddata
\tablecomments{Relevant parameters include; number of particles, N, initial average interparticle distance $\bar{\ell}$, softening parameter, $\epsilon$, minimum scale of turbulent fluctuations $\ell_k$, and initial virial parameter $\alpha_i$. } 
\end{deluxetable}

\subsection{Cluster Finding with FOF}
\par To evaluate the formation of star clusters over time in the simulation, we implement a cluster finder based on the Friends of Friends (FOF) algorithm. FOF has been widely used in the cosmology community as a halo-finder instrumental in extracting  halo mass functions from simulations \citep{knebe11}. The algorithm has only one free parameter, the linking length $b$, and operates on a set of particle positions in three dimensions.
To locate clusters, the algorithm iteratively determines all the particles in a group that are within the linking length $b$ away from at least one other particle in the group. 
\par The FOF algorithm offers a few advantages as a cluster finder: it is easier to capture irregular structures, an advantage when looking for stellar clusters, unlike some density threshold methods that construct halos/clusters out of spheres. With only one free parameter, there is relatively little fine tuning needed to produce mass functions, as long as there is some basis for the selection of the linking length. 
\par To generate complete mass functions that probe substructure across different size scales, we use a hierarchical FOF. This approach utilizes a range of linking lengths to generate several sets of mass functions, with smaller values of the linking length finding smaller sub-structures. The mass functions generated at each scale can be summed, removing duplicate structures where applicable to create one hierarchical mass function. 
We empirically determine the base linking length for our setup by using the value that optimizes both the amount of particles in groups and the number of groups created.  For most of the setups, this value runs close to  $0.2 \bar{\ell}_i $ , where $\bar{\ell}_i$ is the initial average interparticle distance, which guarantees groups that are at least 125 times denser than the initial state and an overdensity of at least 3 above the expected interparticle density if the final state were homogeneous. With hierarchical FOF, we probe down to $0.08 \bar{\ell}_i$ at maximum .

\par The final step for our cluster finder is to filter out structures that are clusters of clusters, which we don't want to include in our analysis. 
In the case of FOF, clusters that are adjacent to one another can be grouped together into one single cluster, even if they are not representative of a single object.  
To filter out these cases, we adopt a criterion based on the behavior of the average cluster density with mass. In a mass accretion scenario that produces large scale structures by the accretion of smaller scale structures onto a larger gravitating bound structure, the mean densities of more massive structures tend to increase. This will not be true for clusters of clusters, which are grouped solely on the basis of their adjacency. The average density of these objects flattens out and decreases, consistent with groups that are growing more in radius than they are in mass. 
 \par
\begin{figure}[h!]
    \centering
    \includegraphics[width=\columnwidth]{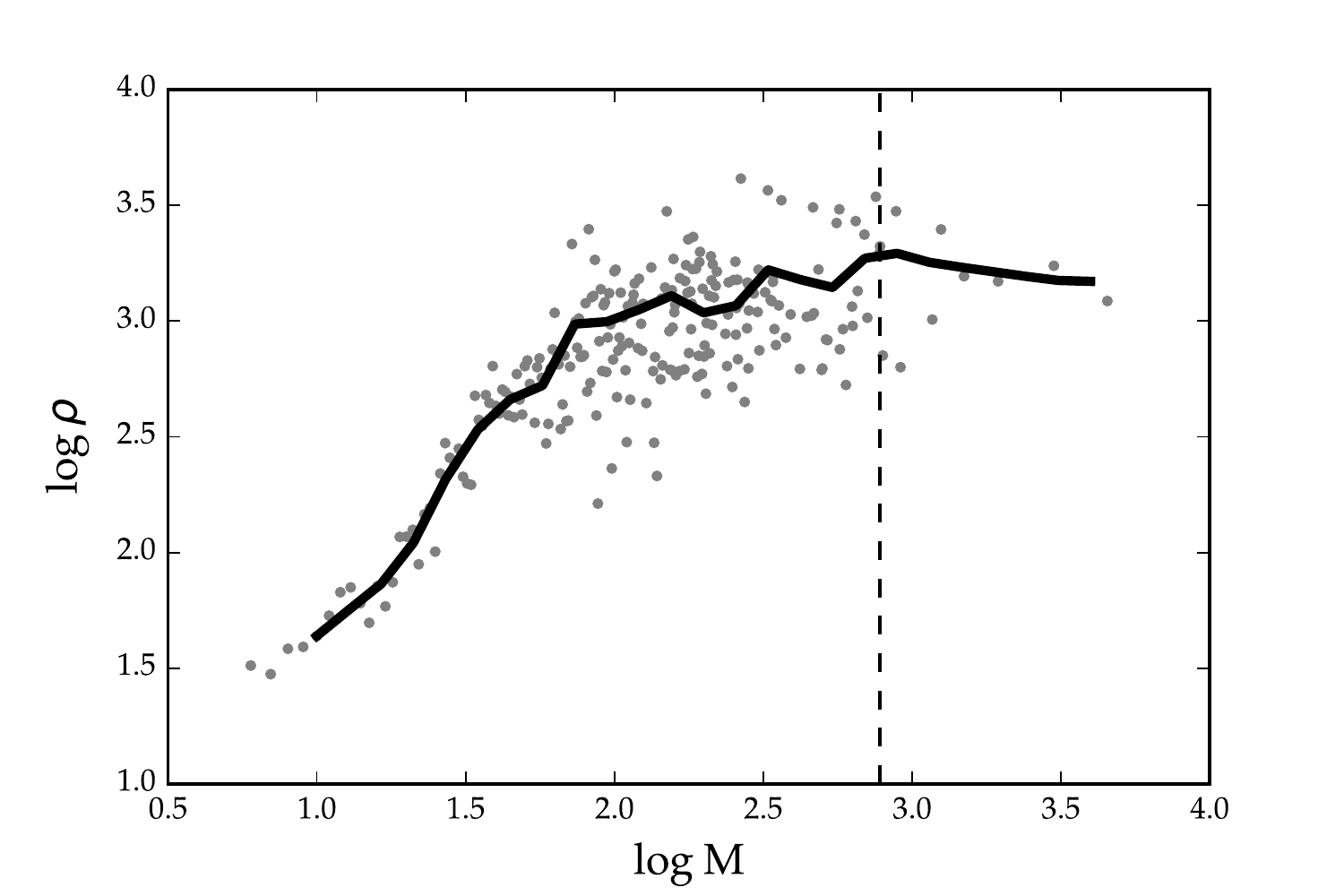}
    \caption{Sample view of the upper mass refinement process. Gray dots show the mean density of groups at that mass. The black solid line is the interpolated mean density binned by mass. The identified threshold location is the maximum of the interpolation, where the overall density trend starts to decrease, shown as a black dashed line. In this example, the mass function will be truncated at masses above $\log M = 2.89$.}
	\label{fig:denref}
\end{figure}

 In Figure \ref{fig:denref}, we see that there exists a rough truncation mass above which the clusters are likely to be clusters of clusters due to a turnover in average cluster density. By only including clusters less massive than the truncation mass in our analysis, we remove clusters of clusters, but keep all sub-structures that comprised them due to the hierarchical nature of the data sets. 

\begin{figure*}[t]
    \centering
    \includegraphics[clip=true,trim = 3cm 0cm 0cm 0cm, width=0.85\paperwidth]{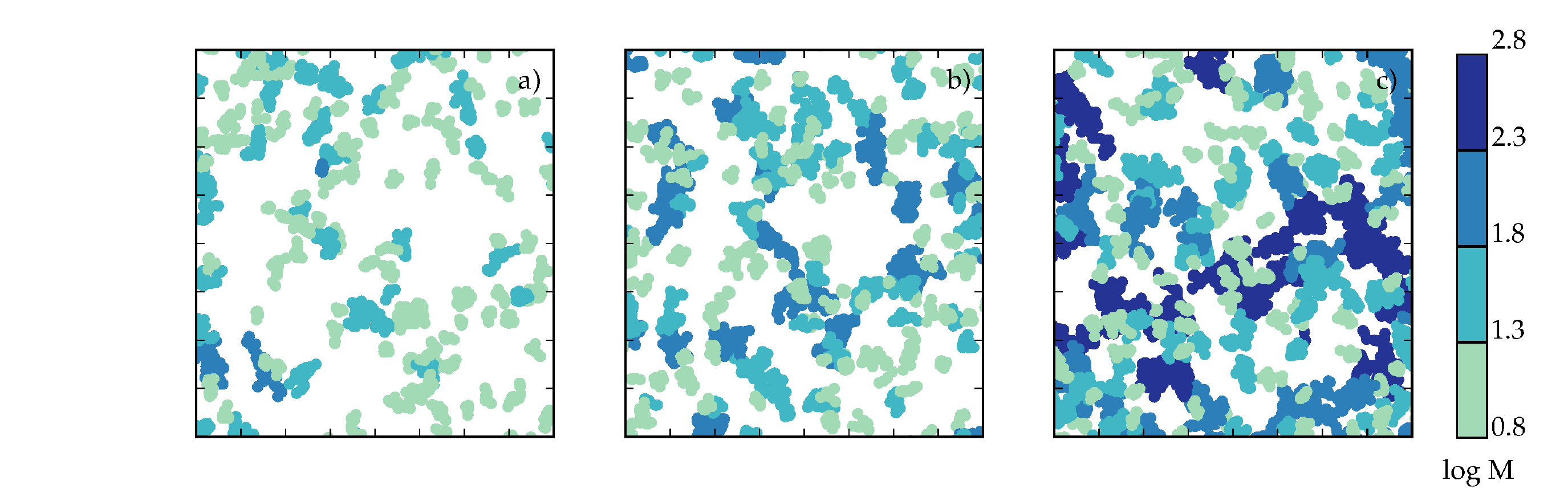}
    \caption{Thick slice of a central box $0.4 R \times 0.4 R \times 0.2 R$ in volume, where clusters are color coded by mass.  In units of initial free-fall time, panels a)-c) are taken at t = 0.8, 1.0, and 1.3. Scaling to typical molecular cloud values, the time between panels is $\Delta t = 2.74 \mathrm{Myr}$.}
    \label{fig:progression}
\end{figure*} 

\subsection{Results}
\subsubsection{Cluster Growth Through Gravitational Focusing}
\begin{figure}[h!]
    \centering
    \includegraphics[width=\columnwidth]{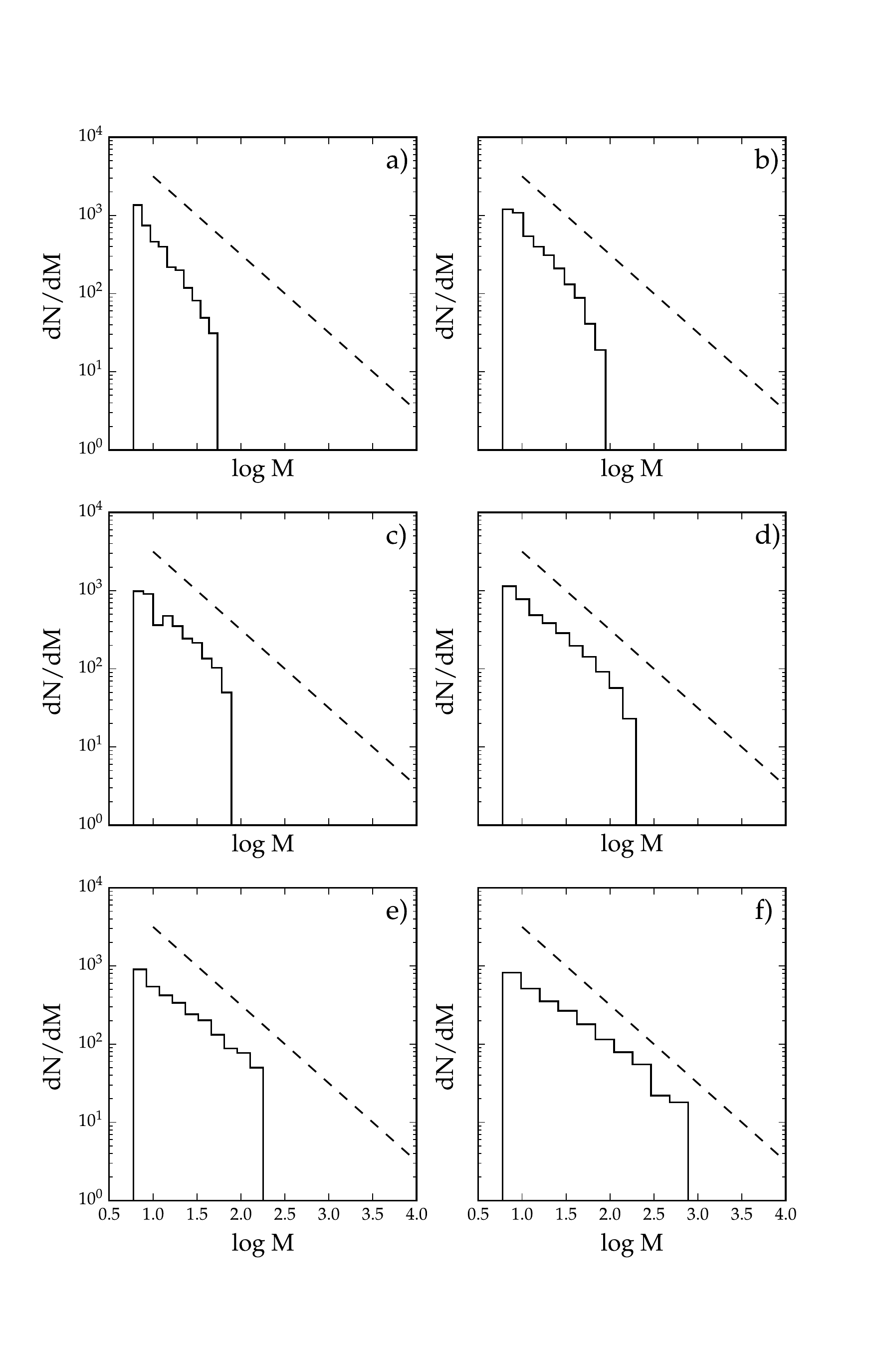}
    \caption{Time evolution of the mass function for run 40s1b, a set of 40000 particles initially distributed in a homogenous spherical distribution. Particle number is used as a proxy for mass here. The dashed line has a slope of $-1$.  Frames a)-f), in units of initial free-fall time, are take at t = 0.52, 0.77, 0.90, 1.02, 1.15, and 1.3 . Adopting typical molecular cloud values we can rescale such that the time between snapshots $\Delta t = 1.37 \mathrm{Myr}$.}
    \label{fig:teval40ksphere1}
\end{figure}

In Figure \ref{fig:progression},  we present a series of three epochs of cluster formation in run 40s1b.
Run 40s1b is an initially spherical distribution of 40000 particles. In Figure \ref{fig:teval40ksphere1}, we show the time evolution of its mass function, where the right panels of the Figure correspond to the three epochs in Figure \ref{fig:progression}. (For the sake of generality, we will use the number of particles in a cluster as a proxy for the mass of the cluster for all figures, i.e. $M = N [\bar{m}]$). Figure \ref{fig:progression} frame a) shows an initial phase where most groups are of a similarly low mass. At this point, over half of the particles in the simulation can be assigned to a group. This distribution is a result of the initial turbulent mixing in the simulation. By the epoch in Figure \ref{fig:progression} frame b), some intermediate mass groups have been created. We see this in the corresponding frame in Figure \ref{fig:teval40ksphere1} d), where the tail of the distribution has started to grow, creating a shallower slope. In the last epoch shown in Figure \ref{fig:progression} c) and the last panel of Figure \ref{fig:teval40ksphere1}, the intermediate mass clusters have rapidly accreted new members, such that many of them are now some of the most massive clusters and the mass function has grown to approach the asymptotic $\Gamma = -1$ slope. While at this stage, all groups found by FOF are bound by virtue of the virial parameter $\alpha < 1$, there is a subset of particles that were not assigned to groups and remain unbound.  We include the final power law fits of the mass function for all runs in Table \ref{tab:runfits}, as well as sample values for fits with small and large numbers of bins. The ``averaged" slope values are a weighted average of fits over a range of bin sizes and different density refinement algorithms.

\begin{figure}[h!]
    \centering
    \includegraphics[width=\columnwidth]{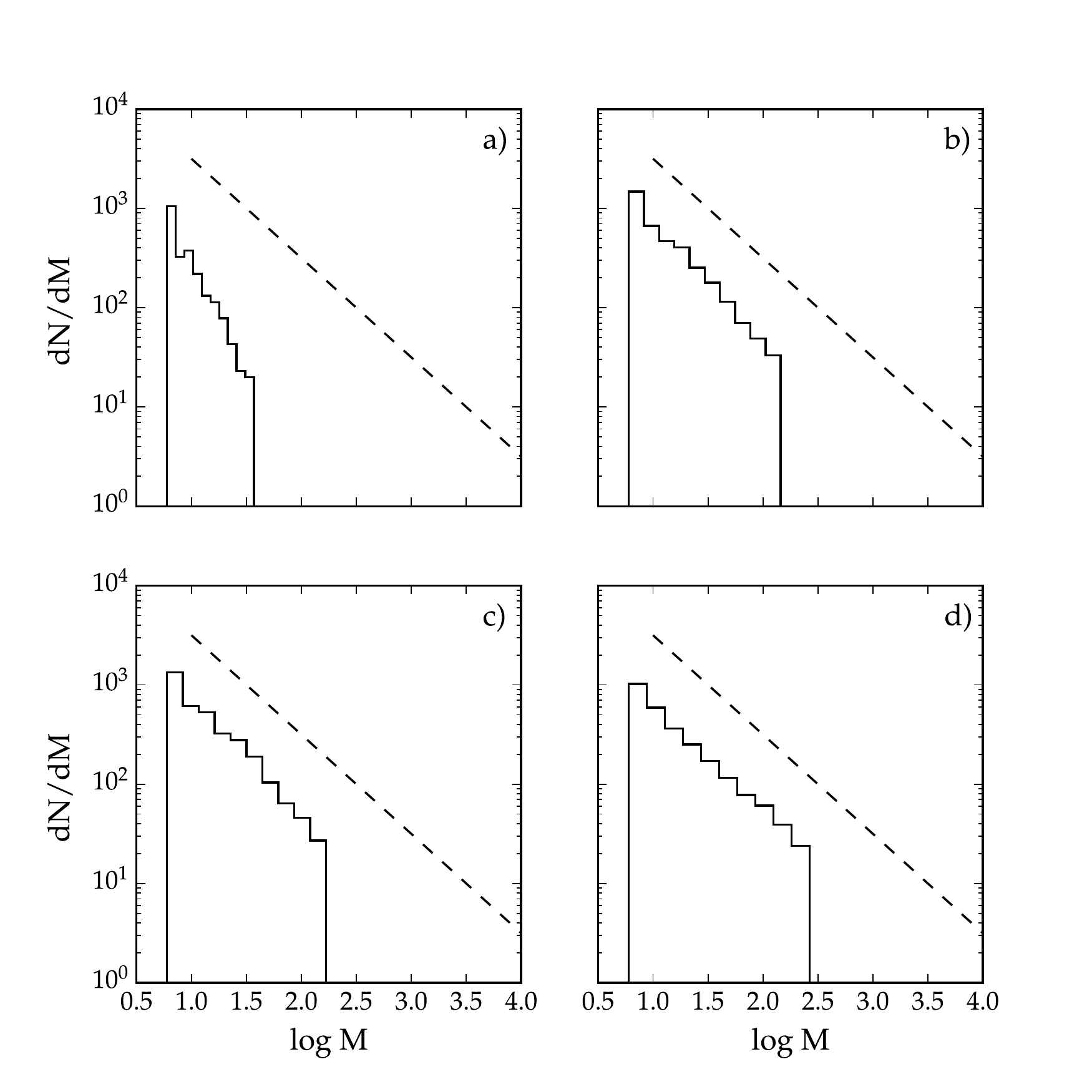}
    \caption{Time evolution of the mass function for run 40d1b, a set of 40000 particles initially distributed in a homogenous disk. Particle number is used as a proxy for mass here. The dashed line has a slope of $-1$.  Frames a)-d), in units of initial freefall time, are taken at t = 0.50, 0.65, 0.80, 0.95}
    \label{fig:teval40kdisk1}
\end{figure}

In the final frame (f), the mass function appears
to be a little flatter than -1 at lower masses, but this is inevitable, the reasons for which are twofold: the initial conditions and the limits on the cluster finder. First, the $\Gamma = -1$ limit is an asymptotic one, and as the starting cloud is finite, eventually smaller mass clumps will simply cease to form due to a lack of replenishing material. Second, while a hierarchical mass function ought to probe the substructure of clumps, we are limited by the smallest linking length we set for the cluster finder: an arbitrary limit due to computational constraints. 
\par In Figure \ref{fig:teval40kdisk1}, we find that the flatter feature does not exist in the disk geometry; the mass functions produce a relatively constant slope across the entire mass range. Disks are prone to an edge effect, where the outside of the disk tends to collect a ring of particles as it contracts \citep{bh04}. This thin layer of excess pile-up creates a region with an overabundance of lower mass groups which, in this case, compensates for the underpopulation of lower mass clusters we see in the spherical geometry. Despite these differences, slopes between the different geometries vary by $<0.2$.

\begin{deluxetable}{ l l r }
\tablecaption{Fits for the mass function slope of all \emph{ChaNGa} runs \label{tab:runfits} }
\tablewidth{0.9\columnwidth}
\tablehead{\colhead{run} & \colhead{binning} & \colhead{slope $\pm$ error}}
\startdata
40s1b  & averaged &   -0.89     $\pm$   0.03     \\
        & min     &   -0.87     $\pm$   0.03     \\
        & max     &   -0.90     $\pm$   0.03     \\ \hline
40s2b  & averaged &   -0.91      $\pm$  0.08     \\ 
       & min      &   -0.80     $\pm$   0.04     \\
       & max      &   -0.95     $\pm$   0.04  \\ \hline
40d1b  & averaged &    -0.99    $\pm$  0.04     \\
       & min      &    -1.07    $\pm$  0.05  \\ 
       & max      &    -1.01    $\pm$  0.04  \\ \hline
40d2b  & averaged &     -1.06   $\pm$  0.03     \\
       & min      &     -1.06   $\pm$  0.03 \\    
       & max      &     -1.09   $\pm$  0.05 \\ \hline
400s1b & averaged &    -0.99     $\pm$  0.05    \\
       & min      &    -0.90    $\pm$   0.03 \\
       & max      &    -1.00    $\pm$   0.02 \\ \hline
400s2b & averaged &    -0.98    $\pm$  0.06     \\
       & min      &    -0.85    $\pm$   0.04  \\
       & max      &     -1.00   $\pm$   0.03 \\ \hline
400d1b & averaged &   -1.18     $\pm$  0.04      \\
       & min      &   -1.20     $\pm$  0.03 \\
       & max      &   -1.16     $\pm$  0.02 \\ \hline
400d2b & averaged &    -1.17    $\pm$  0.04    \\
       & min      &    -1.10    $\pm$  0.03  \\
       & max      &    -1.20    $\pm$  0.03 \\
\enddata
\tablecomments{Slope of the power law mass function is given for each run in terms of 'averaged', min, and max values. For each run, fits were taken using three different density refinement techniques and using 10 different values for the binning.  Min and max values for the fit are sample values taken for the smallest and largest number of bins, respectively. The 'averaged' value corresponds to a weighted average of the fit to the slope across all binnings and truncation masses. }
\end{deluxetable}

\subsubsection{Mass Accretion History}
\par
\begin{figure}[h!]
    \centering
    \includegraphics[width=\columnwidth]{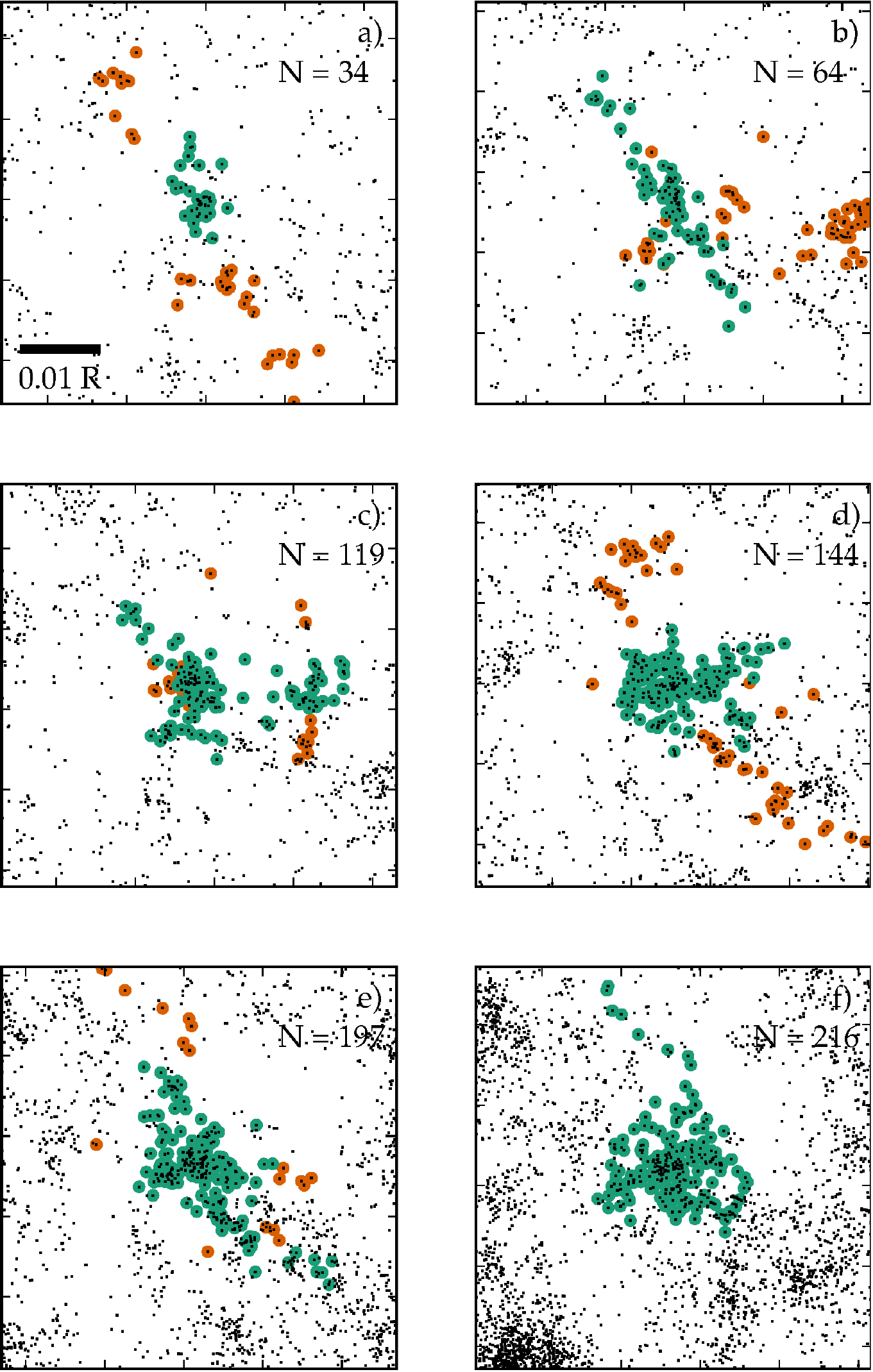}
    \caption{Mass accretion through time for a sample group from run 40s1b on the higher mass tail of the mass function for run 40s1b, each panel corresponds to the one shown in Figure \ref{fig:teval40ksphere1}. Teal circles represent members of the largest group, orange circles represent members that will be accreted onto the group between snapshots. Black dots plot a projection of all particles in view.}
    \label{fig:accretion}
\end{figure}

 For additional insight into the development
of the mass function, we look at the mass accretion history of representative groups. Selecting groups from the power law tail of the distribution, we  characterize the fashion in which the group attains mass by tracing the former groups its members belonged to over time. We can demonstrate that for a typical group at early times there exists a dominant sub-group which will attract smaller groups and accrete them (Figure \ref{fig:accretion}).  Combined with the demonstration that the $\Gamma = -1$ power law is not present at early times, it is unlikely that the mass distribution we see is seeded in the initial conditions, but instead is a function of
gravitationally-driven accretion from a random
distribution.

\begin{figure}[h!]
    \centering
    \includegraphics[width=\columnwidth]{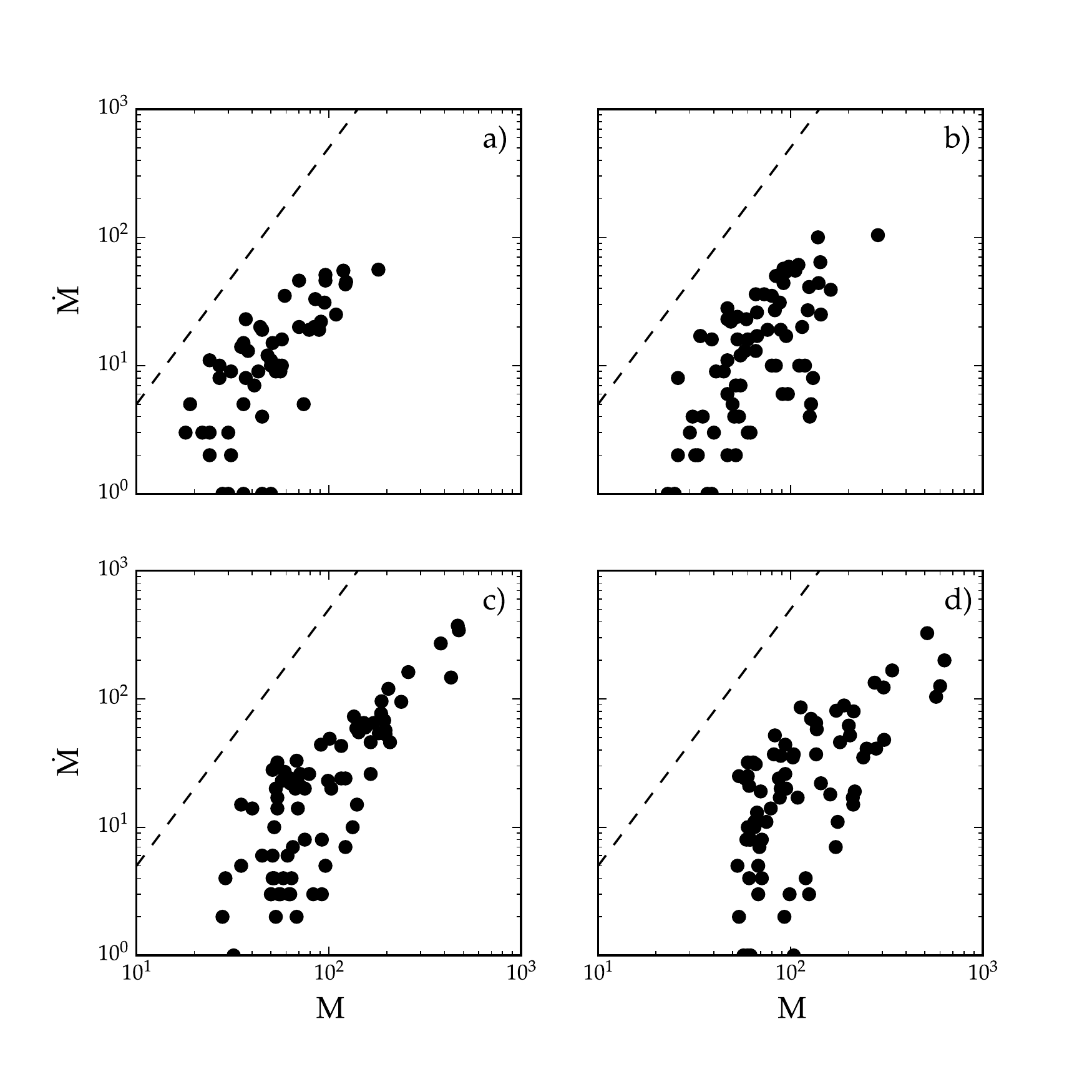}
    \caption{ $M vs \dot{M}$ on an arbitrary scale for run 40s1b. These panels correspond to the last 4 panels of Figures \ref{fig:teval40ksphere1}. The dotted line has a slope of $2$ in log space, representing the canonical $\dot{M} \propto M^2$ relation. }
    \label{fig:mdot}
\end{figure}

\section{Gravitational focusing vs. BHL accretion}

As mentioned in the Introduction, \citet{zinnecker82} showed that if the accretion of an initial population of similar masses scales directly as $M^2$ (as is the limiting case
of BHL accretion), $\Gamma \rightarrow -1$
asymptotically.
 As shown in Fig. \ref{fig:mdot}, $\dot{M}$ vs $M$ starts off as roughly $M^2$, but later shows bending at low and high masses.
The same behavior was seen in BP15, in which the SPH simulation nevertheless produced a $\Gamma = -1$ power law in the sink mass function. BP15 attributed this to “starving” the small masses by competition with higher mass sinks, and slowing down the high masses due to depletion of their environments.
But if the strict BHL accretion formula, or a pure
$\mdot \propto M^2$ is not strictly applicable,
why does $\Gamma \rightarrow -1$?

In the case of the SPH runs with sink formation,
BP15 suggested that the answer was that the
gas densities and velocities were uncorrelated
with the final sink mass; thus on average the
$M^2$ dependence wins out.  They found
that specific {\em local} groups of
sinks, embedded in similar environments, exhibited accretion rates proportional to $M^2$ but with
different values of $\alpha$ (equation \ref{eq:bhl}) for differing groups.  BP15 then argued that
adding up the individual groups
with similar power-law distributions results
in a common power law.

The difficulty with attempting to find an analytic or schematic
understanding of the development of the mass function
is that the environment is not static, but instead is
strongly perturbed by the formation of local centers
of gravitational attraction.  A naive application of the
BHL formula (equation \ref{eq:bhl}) might suggest
that accretion over large enough scales to make massive
clusters is difficult, given the observed increase in
cloud velocity dispersions with increasing size \citep{larson81}.
However, as \citet{heyer09} showed, the Larson velocity-size relation for
molecular clouds is consistent with near-virial motions, as would be expected if gravity were the driving force \citep[see also][]{bp11}.  In this case it is difficult to identify the appropriate value of $v$ to use
in equation \ref{eq:bhl}, and so
formally the BHL accretion picture doesn't apply.  However, the power-law mass function that results and its development with time ( Figures \ref{fig:teval40ksphere1}, \ref{fig:teval40kdisk1}) is quite similar to simulations of pure BHL accretion
\citep[see, e.g., Figure 10 in][]{hsu10}).  Perhaps a schematic way of understanding the results is
that the initial random
velocity fluctuations average out
so that overdensities can accrete
matter at a rate $\propto M^2$.
Whatever the interpretation, the
common results of both the SPH
simulations in \citet{ballesteros15} and the
N-body calculations in this paper
provide strong evidence for
gravitational accretion producing
power law distributions with
$\Gamma \sim -1$.

\section{Discussion}

The simulations in this paper and in \citet{ballesteros15} refer to collections of
particles and sink particles, respectively. Protostar clusters involve a combination of gas and stars.  To apply our results to real
clusters, we must assume 
that the protocluster gas $+$ star mass
function maps directly into the final scIMF.
This seems reasonable because
the total mass in stars is generally thought to
be $\gtrsim 0.3 - 0.5$ of the original protocluster cloud mass
for the cluster to remain gravitationally bound
\citep[unless gas loss is extremely slow;][]
{hills80,mathieu83,geyerburkert01,lada03}. Star formation regions with lower star formation efficiencies will likely not be able to become  clusters at all \citep{kroupa00}. In addition, observations of nearby star clusters, like the ONC, show that clusters themselves form in regions of dense gas at high efficiencies, without significant contributions from diffuse gas components of molecular clouds \citep{hh98}. 
Moreover, the similarity of cluster mass functions
both younger and older than 10 Myr in the interacting galaxy system of the Antennae and
the LMC, suggests that feedback does not alter the shape of the
cluster mass function \citep{fall10}.

Our identification of the cluster mass
function as a result of gravitationally-driven accretion implies that clusters form subvirially.
 Generally speaking, the formation of a massive star cluster must occur in $\lesssim 3-10$~Myr, to avoid disruption by the energy input from the massive stars.  This requires that the free-fall time be $\sim 10$~Myr or less, which requires molecular hydrogen densities $\gtrsim 100 {\mathrm cm^{-3}}$, a plausible value for molecular clouds.  The other requirement is that expansion velocities or dispersions be at most roughly virial (assuming some dissipation of supersonic motions).  While observed velocity dispersions of molecular clouds increase with increasing scales \citep[Larson's first law;][]{larson81}, recent observations suggest that these motions are roughly virial \citep{heyer09}
(see also \citet{larson81}).  If the supersonic velocities are largely gravitationally-driven, avoiding the problem of too rapid dissipation \citep{bp11}, this second criterion is automatically satisfied.

Testing the hypothesis of gravitationally-driven accretion and/or subvirial initial conditions directly with
the currently available observations is difficult \citep{proszkow09,kuznetsova15}, but the advent of Gaia may
make it possible to search for
collapse in statistically-significant
samples of stars, especially combined with radial velocity measurements
\citep{tobin09,kounkel16,dario16}.
The gravitationally-driven accretion seen in simulations tends to form infalling filamentary streams, and this may be observable in the gas using
appropriate molecular tracers.

Morphological tests are possible and even easier
to apply.  As \citet{bh04} showed, gravitational focusing tends to produce
concentrations of mass in finite clouds near regions of smaller radii of curvature at the
cloud edge; the simplest example of this is formation of dense gas near the ends of filamentary
clouds. While this picture is consistent with the spatial structure of the Orion A cloud \citep{hb07},
it might be possible to put this on a firmer statistical basis, using large-scale maps from the {\em Spitzer}
and {\em Herschel} Space Telescopes and other facilities
\citep{churchwell09,andre10,mairs16}.

Our picture of gravitationally-focused accretion naturally explains the tendency of stars to form in groups and clusters.  It is also consistent with the idea that
the supersonic motions in dense molecular clouds tend to be driven by gravity \citep{bp11}.  Finally, this picture suggests a close connection with the upper-mass slope of the stellar IMF \citep{ballesteros15}, with departures from the
limiting $\Gamma = -1$ slope to that of the Salpeter value arguably the result of feedback from high-mass stars halting accretion.

Detailed numerical simulations of star and cluster formation in galaxy models with
complicating effects such as stellar feedback and magnetic fields could probe the
limits of this simple picture and help constrain the expected masses of star clusters as a function
of environment for comparison with observations.

\acknowledgments

AB acknowledges important conversations with Clare Dobbs, LH with
Mike Fall, Javier Ballesteros-Paredes, and Mark Heyer.
This work was supported in part by NASA grants NNX16AB46G and NNH15ZDA001N-XRP,by the University of Michigan, and in part through computational resources and
services provided by Advanced Research Computing at the University of
Michigan, Ann Arbor.

\end{document}